\documentclass[aps, prl, twocolumn, floatfix, superscriptaddress, nofootinbib, nobibnotes]{revtex4-2}
\usepackage{amsmath, amssymb, graphicx, epsfig, xcolor}
\usepackage[colorlinks=true, allcolors=blue]{hyperref}

\begin{document}

\title {Key Role of Charge Disproportionation in Monoclinic
Semiconducting Fe\textsubscript{2}PO\textsubscript{5},\\
a Room-Temperature \emph{d}-Wave Altermagnet Candidate}

\author{Zhen Zhang}
\address{Department of Physics and Astronomy, Iowa State University, Ames, IA 50011, USA}

\author{Mohd Anas}
\address{Department of Chemistry, Biochemistry and Physics,
South Dakota State University, Brookings, SD 57007, USA}

\author{Andrey Kutepov}
\address{Department of Physics and Astronomy and Nebraska
Center for Materials and Nanoscience, University of Nebraska-Lincoln,
Lincoln, NE 68588, USA}

\author{Parashu Kharel}
\address{Department of Chemistry, Biochemistry and Physics,
South Dakota State University, Brookings, SD 57007, USA}

\author{Vladimir Antropov}
\email{antropov@iastate.edu}
\address{Department of Physics and Astronomy, Iowa State University, Ames, IA 50011, USA}
\address{Ames National Laboratory, U.S. Department of Energy, Ames, IA 50011, USA}

\date{March 19, 2026}

\begin{abstract}
\(\beta\)-Fe\textsubscript{2}PO\textsubscript{5} is an emerging room-temperature
\emph{d}-wave altermagnet featuring quasi-one-dimensional crystal and
magnetic structures, orthogonal transport channels for opposite spins,
and large band spin splitting, which is a promising material for
next-generation spintronics and magnonics. However, its
crystal and electronic structures remain inconclusive.
Here, joint experimental and theoretical studies confirm and explain the
appearance of its monoclinic structure and semiconducting band gap. We
discover that an electronic instability appears in the tetragonal
metallic state as the joint effect of density functional theory and
Hubbard U correction (DFT+U) and results in a charge 
disproportionation, which in turn stabilizes the monoclinic distortion
with narrow gap formation. The successful capture of this effect within
DFT+U requires accounting for the relevant symmetry-breaking
energy-lowering channels---charge disproportionation and structural
distortion; otherwise, tetragonal-symmetry-constrained
calculations yield only a metallic state.
Fe\textsubscript{2}PO\textsubscript{5} is thus best described as a
correlation- and hybridization-assisted, distortion-coupled,
charge-disproportionated semiconductor. It represents a rare room-temperature semiconducting \emph{d}-wave altermagnet. It also provides a rare platform for studying the coexistence of altermagnetism and charge density wave in quasi-one-dimensional systems.
\end{abstract}

\maketitle

Antiferromagnets featuring momentum-dependent nonrelativistic spin
splitting, also known as altermagnets, have attracted a lot of attention
recently \cite{1,2,3,4,5,6,7}. Many altermagnets have been identified
theoretically using existing magnetic materials databases \cite{8}.
However, experimental verification of the predicted properties has
progressed far more slowly than theoretical advances. To seek reliable
altermagnets and provide robust guidance for experiments, known
materials with experimentally well-established magnetic properties
should be reanalyzed. Fe\textsubscript{2}PO\textsubscript{5} is such a
system that has recently been proposed to exhibit fascinating orthogonal
transport channels for opposite spins and large spin
splitting \cite{9,10}. However, only its tetragonal phase with metallic
band structure has been studied. Its temperature (\emph{T}) phase
diagram is rather complex, and
the tetragonal phase and metallic state, as we show here, are
irrelevant for room-temperature altermagnetic applications.

\begin{figure}[t]
	\includegraphics[width=\linewidth]{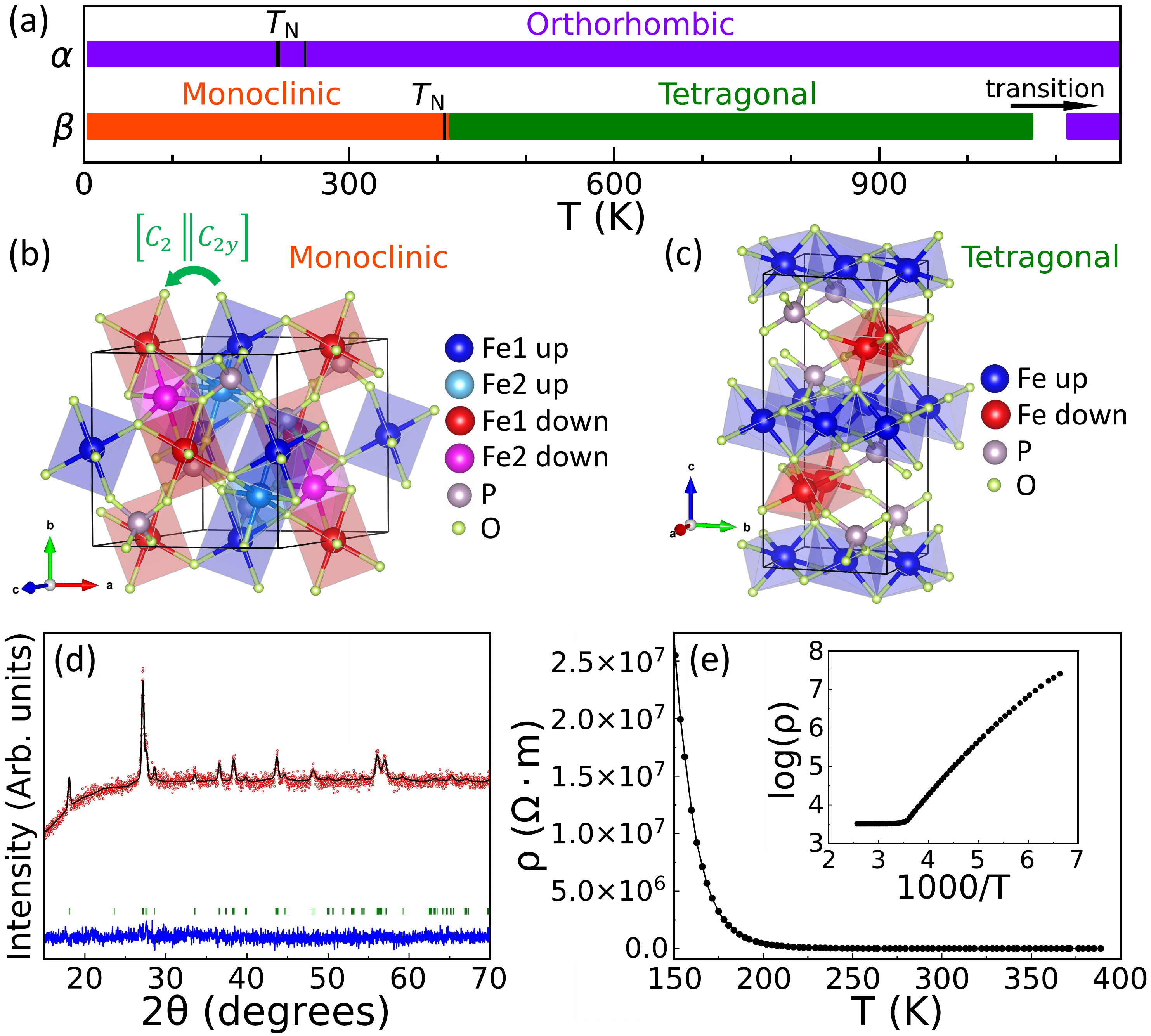}
	\caption{(a) \emph{T} phase diagram of
		Fe\textsubscript{2}PO\textsubscript{5} summarized from literature \cite{11,12,13,14,15,16}.
		Crystal and magnetic structures \cite{14,15,16} of the (b) m-phase and (c) t-phase.
		32-atom conventional cells are shown. (d) The room-temperature XRD
		pattern for \(\beta\)-Fe\textsubscript{2}PO\textsubscript{5} (red), along
		with the Rietveld-simulated pattern for the m-phase (black) and the
		corresponding Bragg positions (green) and difference curve (blue). (e)
		The measured \(\rho(T)\).}
	\label{fig1}
\end{figure}

The \emph{T} phase diagram of polymorphic 
Fe\textsubscript{2}PO\textsubscript{5} summarized from literature \cite{11,12,13,14,15,16} is shown in Fig. \ref{fig1}(a). 
The orthorhombic structure (space group \emph{Pnma}), also known as the
\(\alpha\)-phase, was first observed \cite{11}. Synthesized at high
\emph{T} (1173 K \cite{11}), the \(\alpha\)-phase can be quenched to low
\emph{T} and was measured down to 4.2 K \cite{12}. Its Néel temperature
(\emph{T}\textsubscript{N}) was reported to be 250 \cite{11,13},
220 \cite{11}, or 218 K \cite{12}. Apart from the \(\alpha\)-phase, there is
an isomeric \(\beta\)-phase stabilized in the comparable
\emph{T} range. The tetragonal structure (space group
\emph{I4\textsubscript{1}amd}) was initially observed \cite{14,15}.
Later, the monoclinic structure (space group \emph{I2/a}) was observed
to stabilize as the low-\emph{T} \(\beta\)-phase \cite{16}. The
\(\beta\)-phase overall has a \emph{T}\textsubscript{N} of
408 \cite{15,16} or 410 K \cite{14}, which is very close to its
monoclinic-tetragonal structural transition \emph{T} of 415 K \cite{16}.
At very high \emph{T}, a nonreversible \(\beta\) to \(\alpha\) structural
transition happens at 1073 \cite{15} or 873 K \cite{14} and completes at 1113 K \cite{15}. For brevity, we denote the monoclinic and tetragonal phases as the
m-phase and t-phase, respectively.

The crystal \cite{14,15,16} and magnetic \cite{14,15} structures of the
\(\beta\)-phase have been measured. Although the measurements for the
\(\beta\)-phase's magnetic structure \cite{14,15} were before the
identification of the m-phase \cite{16}, the magnetic structure of the m-phase 
can be confirmed since they were reported by the same group of
authors \cite{14,15,16}. The m-phase and t-phase are displayed in Figs.
\ref{fig1}(b) and \ref{fig1}(c), respectively. Both structures contain chains of
face-sharing FeO\textsubscript{6} octahedra along two orthogonal (for
t-phase) or nearly orthogonal (for m-phase) directions. Fe atoms couple ferromagnetically to each other within each chain. The (nearly)
orthogonal chains couple antiferromagnetically to each other. There are two crystallographically
inequivalent Fe sites alternating along each chain in the m-phase, in contrast to the crystallographically equivalent Fe site in the t-phase. Inequivalent Fe magnetic states were also observed in the
Mössbauer spectra for the \(\beta\)-phase \cite{14}. 
However, by the time this result was reported,
the m-phase was unknown; the authors claimed this result for the
t-phase and did not discuss the nature of the observed magnetic state inequivalence \cite{14}.

In terms of the transport properties, a previous study measured the
\emph{T}-dependent electrical conductivity for the
\(\beta\)-phase, though whether the structure is monoclinic or tetragonal
was not distinguished, and observed a semiconducting behavior with a gap of
0.20 (0.32) eV below (above) 305 K \cite{14}. Previous theoretical
reports on Fe\textsubscript{2}PO\textsubscript{5}
only studied the t-phase and reported it to be a metal, and no attempts
to describe the observed energy gap or Fe magnetic state inequivalence have been made \cite{9,10}.

Here, using electronic structure methods, experimental synthesis, and
measurement for electrical transport properties, we confirm the
\(\beta\)-Fe\textsubscript{2}PO\textsubscript{5} to be a monoclinic and
semiconducting room-temperature \emph{d}-wave altermagnet. The
underlying key role of charge disproportionation (CD), also known as
charge ordering or charge density wave, in the gap formation is discovered and discussed.

To double-check the room-temperature m-structure \cite{16} and the
semiconducting nature \cite{14} of
\(\beta\)-Fe\textsubscript{2}PO\textsubscript{5}, we have synthesized it
(see Methods) and measured its electrical transport properties.
Figure \ref{fig1}(d) shows the X-ray diffraction (XRD) pattern recorded at room
temperature for the \(\beta\)-Fe\textsubscript{2}PO\textsubscript{5}
powder sample. The Rietveld analysis of the XRD pattern indicates that
\(\beta\)-Fe\textsubscript{2}PO\textsubscript{5} crystallizes into a
m-structure (space group \emph{I12/a1}), which is consistent with the
previous report \cite{16}. Using a slightly modified measurement protocol
(two probe method) in the electrical transport option of the
Physical Property Measurement System, we have been able to
measure \emph{T}-dependent resistivity \(\rho(T)\) for
\(\beta\)-Fe\textsubscript{2}PO\textsubscript{5} between 150 and 390 K
(in the \emph{T} range where the crystal structure is monoclinic), which
is shown in Fig. \ref{fig1}(e). \(\beta\)-Fe\textsubscript{2}PO\textsubscript{5}
is highly resistive with a room-temperature resistivity of
3.3×10\textsuperscript{3} \(\Omega\)\(\cdot\)m. The resistivity increases when \emph{T}
decreases below room temperature. Reference \cite{14} has previously
measured the \(\rho(T)\) in the \emph{T} range of 150--570 K. Two
regions were observed in the \(\rho(T)\) curve: 150--305 K and 305--570
K, with the activation energies 0.20 and 0.32 eV, respectively \cite{14}.
The activation energy in the higher-\emph{T} region is greater. At 305
K, a change in the slope of the log(\(\sigma \cdot T\)) vs \emph{T}
curve was seen \cite{14}. This is much lower than the structural
transition \emph{T}, and the change in the log(\(\sigma \cdot T\)) vs
\emph{T} curve cannot be related to the monoclinic to tetragonal
structural transition that occurs at 415 K. Therefore, the m-phase is
semiconducting \cite{14}. We also observed a change in the slope of the
log(\(\rho\)) vs 1000/\emph{T} curve for
\(\beta\)-Fe\textsubscript{2}PO\textsubscript{5}, shown in the inset of
Fig. \ref{fig1}(e), at 285 K, similar to the observation by \cite{14}. Here, the
\(\rho(T)\) curve was fitted using
\(\rho(T) = \rho_{0}{\rm exp}(\frac{E_{a}}{k_{\rm B}T})\), where \(E_{a}\) and
\(k_{\rm B}\) denote activation energy and Boltzmann constant, respectively.
In the lower-\emph{T} region of 150--285 K, \(E_{a}\) was calculated to
be 0.26 eV, which agrees well with \cite{14}. In the higher-\emph{T}
region of 285--390 K, \(E_{a}\) was calculated to be 0.02 eV. The
activation energy in the higher-\emph{T} region is smaller than that
below 285 K. This behavior contrasts with the previous
investigation \cite{14}. Overall, the behavior of the \(\rho(T)\) curve
indicates the semiconducting nature of
\(\beta\)-Fe\textsubscript{2}PO\textsubscript{5} up to 390 K.

\begin{figure}[t]
	\includegraphics[width=\linewidth]{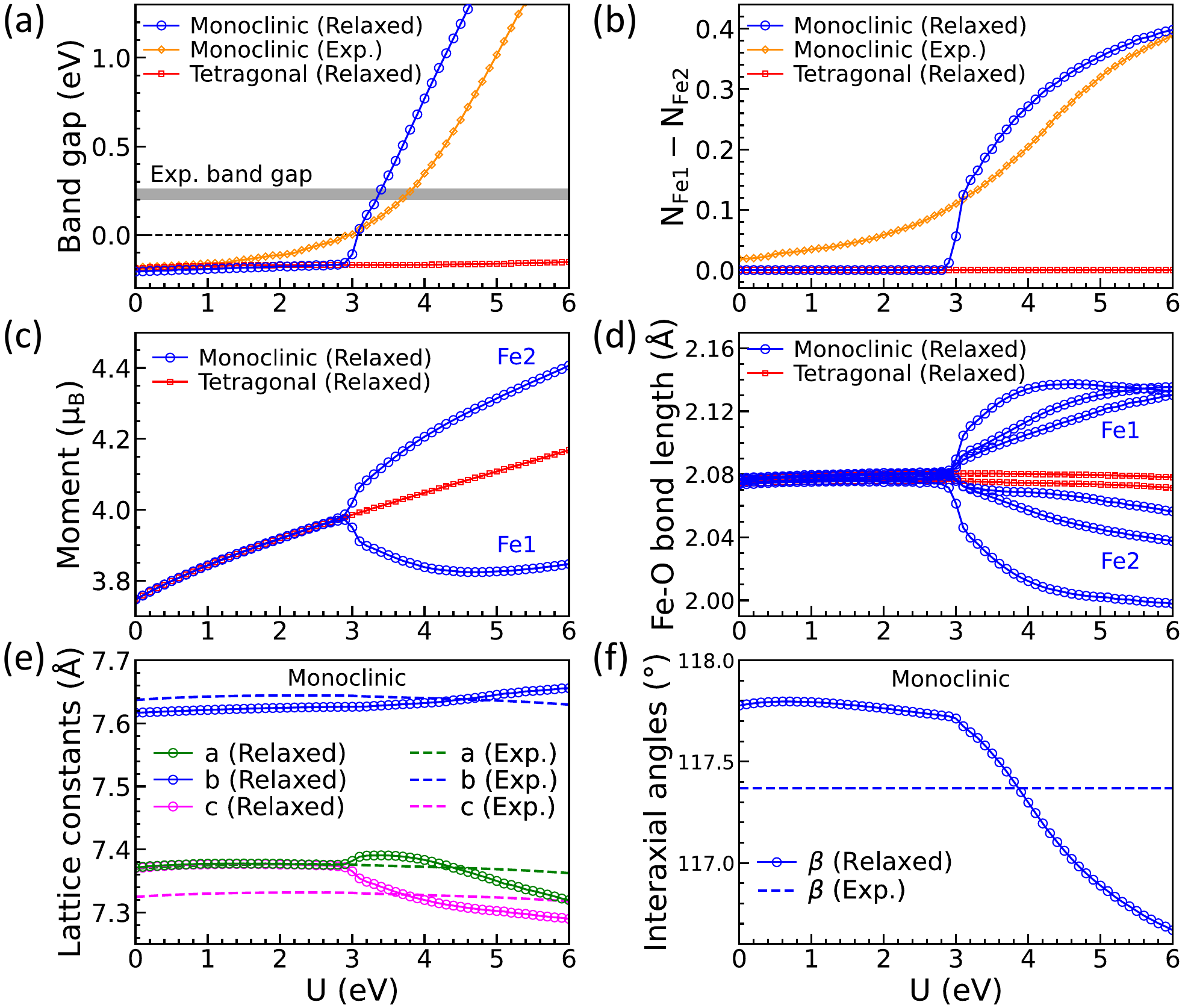}
	\caption{(a) Fundamental band gap \(E_{g}\) calculated in the relaxed
		m-structure, experimental m-structure \cite{16}, and relaxed t-structure.
		In the t-structure, symmetry was turned on, and CD was not allowed. (b)
		Bader charge difference \(\mathrm{\Delta}N\) between the two Fe sites.
		(c) Total magnetic moments on the Fe sites and (d) Fe-O bond lengths in
		the relaxed m- and t-structures. (e) Relaxed (solid curves and dots)
		lattice parameters \emph{a}, \emph{b}, \emph{c}, and (f) \(\beta\)
		compared with the scaled experimental results \cite{16} (dashed lines)
		for the m-structure.}
	\label{fig2}
\end{figure}

The electronic structures of the m- and t-phases were calculated by
density functional theory with a Hubbard U correction (DFT+U) to check
their consistency with the observed semiconducting behavior. For
\(\beta\)-Fe\textsubscript{2}PO\textsubscript{5}, either tetragonal or
monoclinic, a fundamental band gap
(\(E_{g} = E_{\rm CBM} - E_{\rm VBM}\)) can be defined for its semimetal-like or semiconducting
band structure (Figs. S4--S18 of the Supplemental Material \cite{17},
which also includes Refs. \cite{15,16}). The measured semiconducting band
gaps in the lower-\emph{T} region, 0.26 eV from this study and 0.20 eV from \cite{14},
were used for comparison. Firstly, the high-symmetry t-phase was
studied. With symmetry included in the calculations, the charge
difference between Fe sites was not allowed. For U values from 0.0 to
6.0 eV, the t-phase remains a metal with a negative \(E_{g}\) (red curve
in Fig. \ref{fig2}(a)). In contrast, for the m-phase, \(E_{g}\) opens at about U
= 3.0 eV, increases monotonically with increasing U, and reaches the
experimental gap at about U = 3.4 eV (blue curve in Fig. \ref{fig2}(a)). A Bader CD
between the two inequivalent Fe sites
(\(\mathrm{\Delta}N = N_{\rm Fe1} - N_{\rm Fe2}\)) also develops above about U =
3.0 eV (blue curve in Fig. \ref{fig2}(b)). In the above calculations, the
structure was relaxed by DFT+U. The abrupt changes in \(E_{g}\) and
\(\mathrm{\Delta}N\) at U = 3.0 eV (blue curve in Figs. \ref{fig2}(a) and \ref{fig2}(b)),
as we show below, are associated with a substantial monoclinic distortion (m-distortion)
developed above this threshold U. For comparison, we also calculated for
the fixed, experimental m-structure \cite{16}. Since the m-distortion in
the experimental structure is already substantial, \(E_{g}\) and
\(\mathrm{\Delta}N\) develop smoothly (orange curve in Figs. \ref{fig2}(a) and
\ref{fig2}(b)). In this case, \(E_{g}\) reaches the experimental gap at about U =
3.8 eV. Interestingly, in both the relaxed and the experimental
m-structures, our calculations give a nearly identical
\(\mathrm{\Delta}N\) = 0.19 at their respective required U for the
experimental gap. A magnetic state inequivalence, indicated by two different magnetic
moments of Fe, also develops in the m-phase above about U = 3.0 eV (blue
curves in Fig. \ref{fig2}(c)). At U = 3.4 eV required for the band gap, the
moment difference is 0.24 \(\mu_{B}\). In contrast, this
difference is absent in the t-phase.

To understand how m-distortion develops, Fe-O bond lengths are compared
between the relaxed m- and t-structures. Fe-O bond lengths for the two
inequivalent Fe atoms in the m-phase split from the tetragonal
counterparts above about U = 3.0 eV, forming a breathing distortion of
FeO\textsubscript{6} octahedra (Fig. \ref{fig2}(d)). (Note that in the m-phase,
there are three pairs of Fe-O bond lengths for each Fe, with each pair
having the same value. In the t-phase, there are four bonds with one
length and two bonds with another.) The m-distortion is also indicated
by the broken symmetry in lattice vector \emph{a} = \emph{c} (Fig. \ref{fig2}(e))
and the abrupt change in monoclinic angle \(\beta\) (Fig. \ref{fig2}(f)) at about
U = 3.0 eV. (Note that \emph{a} = \emph{c} is another indicator of the
t-structure. The tetragonal cell can be transformed to the monoclinic
cell by the relation provided in \cite{16}, and vice versa.) We further
compared these relaxed monoclinic lattice parameters with the
experimental ones \cite{16}. Since relaxed and experimental cell volumes
are not precisely the same, to enable direct comparisons, we scaled the
experimental volume to be identical with the relaxed volume at each U,
with the cell shape unchanged. Calculations reproduce
the experimental \emph{a}, \emph{b}, \emph{c}, and \(\beta\) (dashed
lines in Figs. \ref{fig2}(e) and \ref{fig2}(f)) at U = 3.7--4.5 eV, which are generally
consistent with yet slightly larger than the required U = 3.4 eV for the
band gap. Results in Fig. \ref{fig2} manifest that the relaxed m-phase
converges to the t-phase at about U \textless{} 3.0 eV, and the
m-distortion only becomes substantial at U \textgreater{} 3.0 eV.

We have shown that self-consistent DFT+U can reproduce the experimental
band gap and structural distortion. But the gap formation mechanism is
still unclear. We attempted to decouple the influences of the two
factors, Hubbard U and structural distortion, when describing this gap
formation. The m-phase is
characterized by breathing distortion, i.e., the alternation of bond
lengths between the neighboring FeO\textsubscript{6} octahedra along
each chain. Here, to facilitate the expression for the magnitude of the
distortion, we define the degree of bond alternation (DBA) \cite{18} as
\({\rm DBA} = 2\left( \overline{d{\rm _{Fe1 - O}^{long}}} - \overline{d{\rm _{Fe2 - O}^{short}}} \right)/\left( \overline{d{\rm _{Fe1 - O}^{long}}} + \overline{d{\rm _{Fe2 - O}^{short}}} \right)\).
To ensure the distortion corresponding to each DBA is realistic,
m-structures relaxed at various U in Fig. \ref{fig2}(d) were taken.

\begin{figure}[t]
	\includegraphics[width=\linewidth]{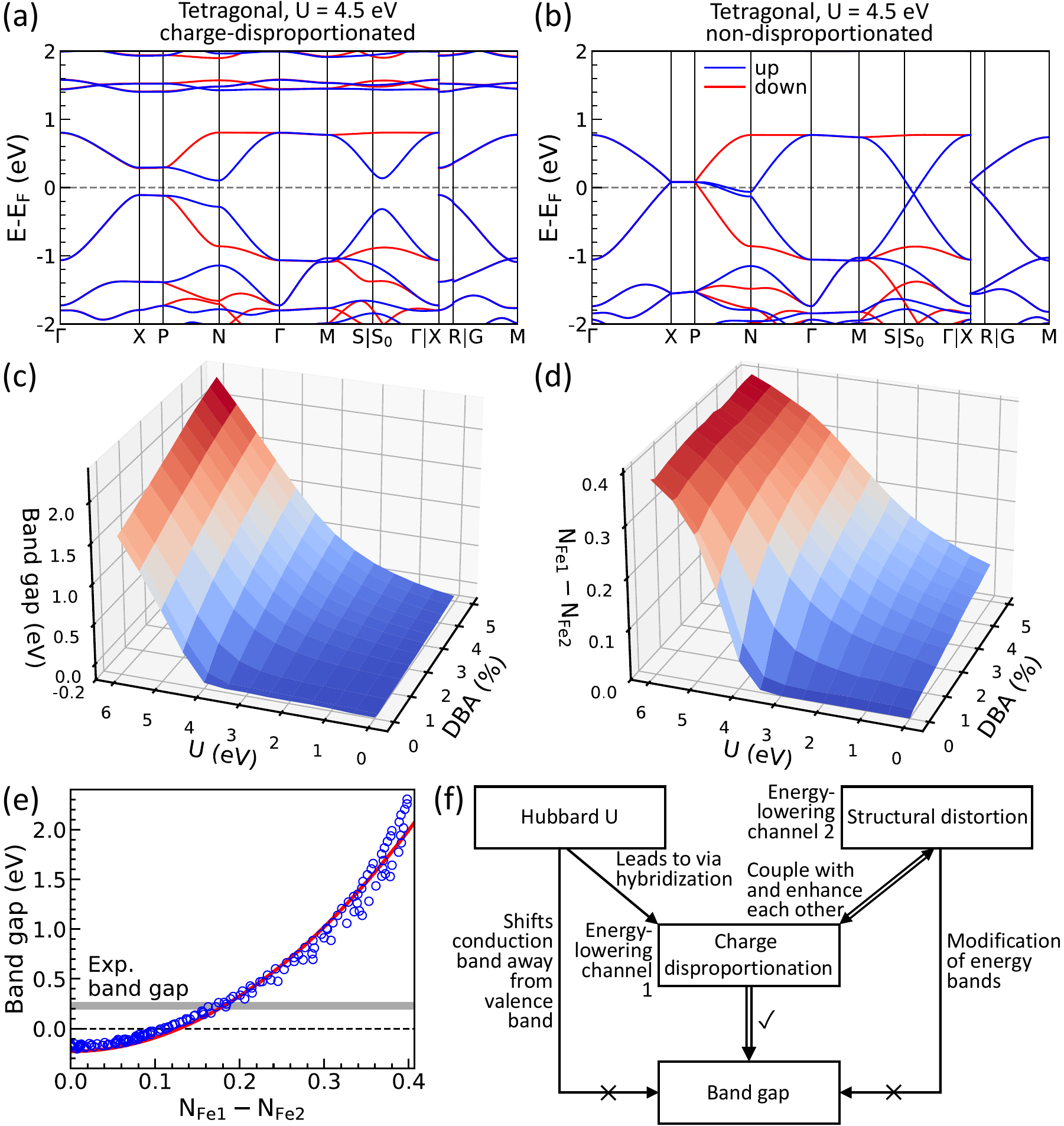}
	\caption{Band structures of the t-phase calculated with U = 4.5 eV in the
		 (a) symmetry-unconstrained charge-disproportionated \cite{20} and 
		 (b) symmetry-constrained non-disproportionated cases. (a) and
		(b) have the same t-structure relaxed by a symmetry-constrained
		calculation at U = 4.5 eV. (c) \(E_{g}\) and (d) Bader
		\(\mathrm{\Delta}N\) calculated with different U and DBA. 
		For DBA = 0, symmetry was turned off, and CD was allowed. (e) The
		\(E_{g}\) vs \(\mathrm{\Delta}N\) relation from (c)(d). (f) Schematic
		interplay among several key physical quantities.}
	\label{fig3}
\end{figure}

In addition, we have also reanalyzed DBA = 0, which corresponds to the
ideal t-structure without m-distortion. In this case, by turning off
symmetry in electronic structure calculations and applying a
symmetry-breaking perturbation to the two Fe sites' occupation
matrix \cite{19}, a CD similar to that in the m-structure can be induced
in the t-structure at about U \textgreater{} 3.0 eV. For example, at U =
4.5 eV, following the electronic variational self-consistency, the
t-phase is trapped in a charge-disproportionated state that is lower in
energy by 6.3 meV/f.u. compared to the symmetry-constrained
non-disproportionated state with the same U. At this state, \(\mathrm{\Delta}N\) = 0.18
and \(E_{g}\) = 0.22 eV (Fig. \ref{fig3}(a)), in contrast to the gapless
non-disproportionated state (Fig. \ref{fig3}(b)). Upon CD, the two pairs of bands near the Fermi level exhibit bandwidth reduction. Further, by breaking the
crystallographically tetragonal symmetry, following geometric
relaxation with the same U, the m-phase is stabilized and even lower in energy by 88.9
meV/f.u compared to the tetragonal charge-disproportionated phase. Such
a comparison explains how electronic and structural symmetry-breaking
energy-lowering channels stabilize the final monoclinic
charge-disproportionated phase.

\begin{figure}[t]
	\includegraphics[width=\linewidth]{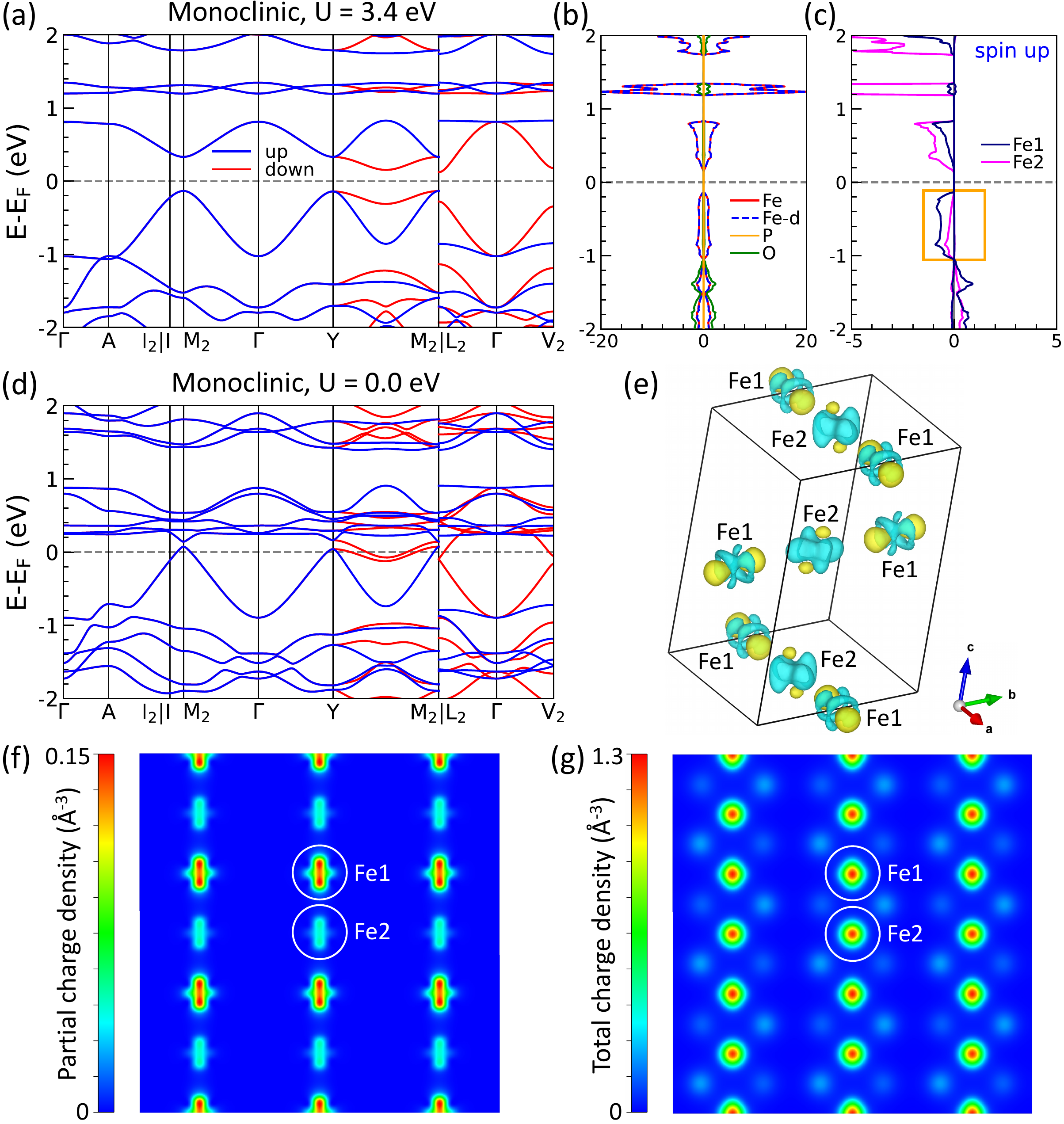}
	\caption{Band structures of the m-phase calculated with (a) U = 3.4
		eV and (d) U = 0.0 eV. (a) and (d) have the same m-structure relaxed at
		U = 3.4 eV. (b) PDOS of each element in one primitive cell 
		(in eV\textsuperscript{-1} spin\textsuperscript{-1} cell\textsuperscript{-1}). (1 cell = 2 f.u.)
		(c) PDOS of one spin-up Fe1 atom and one spin-up Fe2 atom 
		(in eV\textsuperscript{-1} spin\textsuperscript{-1} atom\textsuperscript{-1}). 
		(e) Total charge density difference between (a) and (d). Blue and
		yellow correspond to -0.005 and 0.005 \AA\textsuperscript{-3} isosurfaces,
		respectively. (f) Partial charge density in the \emph{ab}-plane for electronic states located
		within the yellow window in (c). (g) Total charge density in the
		\emph{ab}-plane. (b)(c)(f)(g) all correspond to (a).}
	\label{fig4}
\end{figure}

Figures \ref{fig3}(c) and \ref{fig3}(d) show \(E_{g}\) and \(\mathrm{\Delta}N\),
respectively, obtained with various U and DBA, including DBA = 0. Both
\(E_{g}\) and \(\mathrm{\Delta}N\) are positively correlated with both U
and DBA. As a result, a series of (U, DBA) combinations can reproduce
the experimental band gap. Across the set of calculations shown in Figs.
\ref{fig3}(c) and \ref{fig3}(d), there is an almost one-to-one mapping between \(E_{g}\)
and \(\mathrm{\Delta}N\) (Fig. \ref{fig3}(e)), indicating that the \(E_{g}\) can
be uniquely determined by the Bader \(\mathrm{\Delta}N\), and vice
versa. A positive \(E_{g}\) opens at about \(\mathrm{\Delta}N\) = 0.12,
indicating that states with smaller \(\mathrm{\Delta}N\) remain metallic. It
is followed by a convex increase for larger \(\mathrm{\Delta}N\), and
\(E_{g}\) reaches the experimental gap at about \(\mathrm{\Delta}N\) =
0.19, which is in excellent agreement with the results we obtained
previously either in relaxed or experimental m-structures (Figs. \ref{fig2}(a) and
\ref{fig2}(b)). Overall, the \(E_{g}(\mathrm{\Delta}N)\) can be
phenomenologically fit to a quadratic relation,
\(E_{g}(\mathrm{\Delta}N) = a + b{(\mathrm{\Delta}N)}^{2}\), which is
indicated by the red curve in Fig. \ref{fig3}(e). \(a\) and \(b\) are
obtained as -0.23 eV and 13.89 eV, respectively. Different choices of
crystal structures (e.g., distorted or undistorted, experimental or
relaxed using different U values) in turn require
different U values to reproduce the experimental gap. Our
findings indicate that these choices do not change the resulting \(\mathrm{\Delta}N\).

The concluded electronic structure of monoclinic
Fe\textsubscript{2}PO\textsubscript{5} reproducing the experimental band
gap at U = 3.4 eV is shown in Fig. \ref{fig4}(a). The top of the valence band and
the bottom of the conduction band are formed by a pair of bands,
respectively, which are dominated by the Fe-\emph{d} orbitals (Fig. \ref{fig4}(b)).
These two pairs of bands are semimetal-like when the CD is zero or small
(Figs. S4--S18 \cite{17}). Below -1.06 eV, contributions from O-p
orbitals become important as well. The Fe-O hybridization is indicated
by the coincident peaks in Fe's and O's partial density of states (PDOS)
(Fig. \ref{fig4}(b)). Such a hybridization goes to deeper energies of the valence
band (Fig. S3 \cite{17}). In addition, the short Fe-Fe distance (2.698 \AA)
along each chain of face-sharing FeO\textsubscript{6} octahedra also
facilitates direct Fe-Fe hybridization.

The two Fe sites exhibit different PDOS (Fig. \ref{fig4}(c)) and charge density
distribution. The latter is clearly visualized in the partial charge
density map (Fig. \ref{fig4}(f)) obtained for electronic states located within
-1.06 eV from the Fermi level (yellow window in Fig. \ref{fig4}(c)). The
difference in the two Fe's total charge density is less apparent, yet
still discernible: Fe1 with more charge has a less spherical
distribution than that of Fe2 (Fig. \ref{fig4}(g)). Figure \ref{fig4}(e) further
demonstrates how charge moves upon CD. The state before CD is
approximated by a calculation in the same m-structure, but U = 0.0 eV.
The electronic structure of this reference state is gapless and shown in
Fig. \ref{fig4}(d) for comparison. Although the symmetry-broken m-structure
inevitably includes some CD, the magnitude is rather small at 0.04, and
our demonstration is mostly valid. Upon CD, the charge moves from the
blue regions to the yellow regions in Fig. \ref{fig4}(e).

Fe\textsubscript{2}PO\textsubscript{5} appears to be like the rare-earth
nickelates YNiO\textsubscript{3} and LuNiO\textsubscript{3}, yet with
some differences. YNiO\textsubscript{3} and LuNiO\textsubscript{3}
exhibit analogous alternating breathing distortion of
NiO\textsubscript{6} octahedra in their insulating monoclinic phase, for
which a CD mechanism was proposed \cite{21}. However, later, an absence
of CD with a negligible \(\mathrm{\Delta}N\) between the two Ni sites
(d-shell \(\mathrm{\Delta}N\) of 0.01 in DFT and 0.02 in DFT+U) was
reported \cite{22}. This discrepancy between the CD effect and the
negligible \(\mathrm{\Delta}N\) can be reconciled by a charge
self-regulation mechanism \cite{23,24} between Ni-O. In contrast, here
for Fe\textsubscript{2}PO\textsubscript{5}, the charge
variation (Bader \(\mathrm{\Delta}N\) of 0.19 and d-shell
\(\mathrm{\Delta}N\) of 0.15) is substantial.

Here, DFT over-delocalizes Fe-\emph{d} electrons in
Fe\textsubscript{2}PO\textsubscript{5} and leads to a false metallic
state. Introducing a Hubbard U enhances Fe-\emph{d} localization. The on-site
Hubbard U and inter-site hybridization together stabilize a
charge-disproportionated solution and an opened gap consistent with
experiment. Notably, even in the high-symmetry t-structure where all Fe
sites are crystallographically equivalent, symmetry-unconstrained DFT+U
converges to a lower-symmetry charge-disproportionated insulating state,
indicating that the CD and band gap are not merely a trivial consequence
of structural inequivalence but reflect an intrinsic correlation- and
hybridization-assisted electronic instability.

A schematic diagram depicting the complex interplay among Hubbard U,
structural distortion, CD, and band gap in this system is shown in Fig.
\ref{fig3}(f). In general, the Hubbard U in DFT+U shifts energies of
Fe-\emph{d}-orbitals-containing valence and conduction bands away from each
other. However, here, under tetragonal symmetry protection and without
CD, this effect cannot directly open a gap between those two pairs of
bands near the Fermi level that are dominated by Fe-\emph{d} orbitals. By
properly accounting for the relevant symmetry-breaking energy-lowering
channels \cite{23}---CD and structural distortion in this case---the
observed small band gap can be captured within DFT+U, without going
beyond such a band structure approach. Previous studies \cite{9,10}
reporting an incorrect metallic band structure of
Fe\textsubscript{2}PO\textsubscript{5} by DFT+U are most likely due to
their reliance on commonly employed symmetry-constrained calculations
in the high-symmetry t-phase. On the
other hand, structural distortion can open a band gap via modification
of energy bands. However, the mere removal of the band degeneracy and
the failure to open a positive/global gap (Fig. \ref{fig4}(d)) at large DBA and
zero or small U (Fig. \ref{fig3}(c)) indicates that the gap does not emerge from
a mainly structural distortion effect. Nevertheless, structural
distortion can couple with CD and enhance each other. Therefore,
Fe\textsubscript{2}PO\textsubscript{5} is best described as a
correlation- and hybridization-assisted, distortion-coupled,
charge-disproportionated semiconductor.

Fe\textsubscript{2}PO\textsubscript{5} simultaneously exhibits spin
splitting of electronic bands (e.g., along Y-M\textsubscript{2},
L\textsubscript{2}-\(\Gamma\)-V\textsubscript{2} in Fig. \ref{fig4}(a)) and zero
net magnetization. According to \cite{1}, it is classified as a 
\emph{d}-wave altermagnet. So far, experimentally confirmed
\emph{d}-wave altermagnets include the tetragonal oxychalcogenides family \cite{25,26} and Mn\textsubscript{5}Si\textsubscript{3} \cite{27,28,29}, all of which are metallic. While the magnetic ground state of the first proposed
\emph{d}-wave altermagnet, RuO\textsubscript{2}, is still
controversial \cite{30,31,32}. In altermagnets, opposite-spin sublattices
can be transformed into each other by rotational or mirror symmetries.
Two such transformations within nonrelativistic spin groups are
\(\left\lbrack C_{2} | C_{2y} \right\rbrack\) (Fig. \ref{fig1}(b))
and \(\left\lbrack C_{2} | M_{y}t \right\rbrack\). The spin splitting reaches up to 0.6 eV near the Fermi level, which is on the same order as the large spin splittings of about 1 eV reported for MnTe and CrSb \cite{2}. There is
also lifted degeneracy for Weyl nodal lines and points protected by the
tetragonal symmetry near the Fermi level \cite{9,33}. One such nodal
line, e.g., is along the corresponding k-path in the t-phase to the
Y-M\textsubscript{2} path here in the m-phase (Fig. \ref{fig4}(a)). Upon CD and
m-distortion, the nodal line splits into the two red bands along
Y-M\textsubscript{2} (Fig. \ref{fig4}(a)), which are separated by the gap. 

\begin{figure}[t]
	\includegraphics[width=\linewidth]{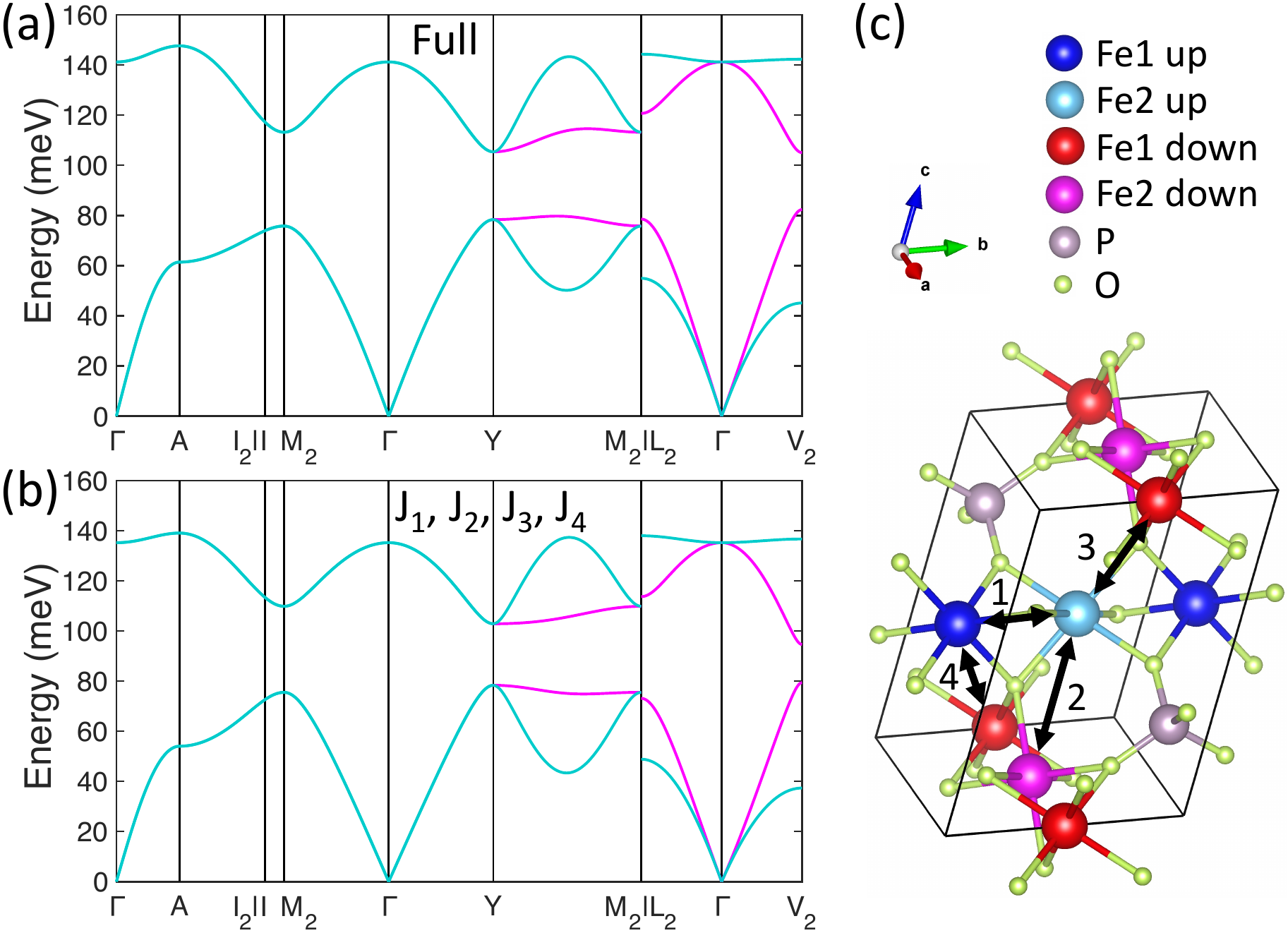}
	\caption{(a) Magnonic band structure of the m-phase calculated with
		U = 3.4 eV. Cyan and magenta indicate opposite chiralities. (b) Magnonic
		band structure obtained from the first four exchange coupling
		parameters, \emph{J\textsubscript{1}}, \emph{J\textsubscript{2}},
		\emph{J\textsubscript{3}}, and \emph{J\textsubscript{4}}. (c)
		Double-headed arrows show the first four nearest Fe-Fe neighbors in the
		16-atom primitive cell of the m-phase.}
	\label{fig5}
\end{figure}

In altermagnets, similar to the electronic bands,
magnonic bands are also expected to exhibit band chiral splitting, with
compensation for bands of opposite chiralities in the Brillouin
zone \cite{34,35,36,37,38}. The calculated magnonic bands of
Fe\textsubscript{2}PO\textsubscript{5} are shown in Fig. \ref{fig5}(a). They
exhibit large chiral splitting (e.g., along Y-M\textsubscript{2},
L\textsubscript{2}-\(\Gamma\)-V\textsubscript{2}), which reaches up to 40
meV. Note that chiral magnon splittings have been reported
for RuO\textsubscript{2} (10 meV) \cite{34}, MnTe (2 meV) \cite{35}, and
CrSb (10--30 meV) \cite{36,37}. The dispersion is linear at the zone
center, which is similar to that for conventional antiferromagnets. A
global magnonic band gap of 20 meV is present between the two lower and
two upper bands. The full magnonic bands (Fig. \ref{fig5}(a)) can be well
approximated by the magnonic bands (Fig. \ref{fig5}(b)) calculated from the first
four magnetic exchange couplings (Fig. \ref{fig5}(c)), \emph{J\textsubscript{1}},
\emph{J\textsubscript{2}}, \emph{J\textsubscript{3}}, and
\emph{J\textsubscript{4}} = 19.1, -24.0, -13.2, and -6.2 meV,
respectively. The intrachain ferromagnetic \emph{J\textsubscript{1}} and
interchain antiferromagnetic
\emph{J\textsubscript{2}--J\textsubscript{4}} dominate the magnetic
interactions in Fe\textsubscript{2}PO\textsubscript{5}. The energy range
of the magnonic dispersion is within 150 meV, which can be verified by
future neutron experiments.

In summary, the appearance of monoclinic distortion and
semiconducting gap in \(\beta\)-Fe\textsubscript{2}PO\textsubscript{5} is confirmed
experimentally and explained theoretically. By using detailed electronic
structure studies, we predict the existence of a charge
disproportionation effect in this system and demonstrate its key role in
the observed narrow gap formation and structural distortion. An
electronic instability appears in the tetragonal metallic state as the
joint effect of on-site Hubbard correlation and inter-site hybridization
and results in the charge disproportionation, which in turn
stabilizes the monoclinic distortion with narrow gap formation. To the best of our knowledge, Fe\textsubscript{2}PO\textsubscript{5} represents the first room-temperature semiconducting \emph{d}-wave altermagnet, while also exhibiting large spin splitting of electronic bands and large chiral splitting of magnonic bands. It also provides a rare platform for studying the coexistence of altermagnetism and charge density wave in quasi-one-dimensional systems. More detailed experiments verifying our predictions in this exciting material are motivated.\\

\noindent \textbf{Methods}

\emph{Sample synthesis and characterization}---The
\(\beta\)-Fe\textsubscript{2}PO\textsubscript{5} was synthesized using
(NH\textsubscript{4})\textsubscript{2}HPO\textsubscript{4} and
Fe(NO\textsubscript{3})\textsubscript{3}\(\cdot\)9H\textsubscript{2}O salts. A
measured amount of these salts was dissolved in distilled water, and the
solution was allowed to evaporate until a dry product was obtained. The
resulting product was allowed to decompose by heating at 400 °C for 9
hours in a continuous flow of nitrogen gas. Later, the powder was
treated with a reducing mixture of argon/hydrogen (10\% hydrogen) gas at
450 °C for 12 hours, following a procedure mentioned in \cite{15}. The
obtained powder was pressed into pellets, sealed in quartz tubes
partially filled with argon, and sintered at 550 °C for 12 hours. The
crystal structure and phase of the prepared materials were investigated
by X-ray diffraction (XRD) measurements using the Rigaku MiniFlex600
diffractometer employing Cu K\(\alpha\) radiation. The obtained lattice
constants of monoclinic \(\beta\)-Fe\textsubscript{2}PO\textsubscript{5}
from Rietveld analysis, \emph{a} = 7.3101(44), \emph{b} = 7.5368(33),
and \emph{c} = 7.2925(39) \AA, are close to those reported
previously \cite{16}. The monoclinic angle \(\beta\) was kept constant at
117.368° as used in the previous report while carrying out the
refinement \cite{16}. We also identified very weak peaks in the XRD
pattern from FePO\textsubscript{4} impurity. Furthermore, the
investigation of thermomagnetic curves \emph{M(T)}, as presented in Text
S1 and Fig. S1 \cite{17}, reveals a magnetic transition near the
previously reported Néel temperature for the monoclinic phase, along
with indications of a small amount of ferromagnetic impurities. However,
such ferromagnetic impurities could not be identified from the analysis
of the XRD pattern.

\emph{Computational methods}---We conducted DFT+U calculations with the
Perdew-Burke-Ernzerhof (PBE) \cite{39} generalized gradient approximation
(GGA) functional using the VASP package \cite{40}. A plane-wave basis set
with a kinetic energy cutoff of 600 eV was used. A Gaussian smearing of
0.05 eV was used. The convergence thresholds were 10\textsuperscript{-5}
eV for electronic self-consistency and 0.001 eV \AA\textsuperscript{-1}
for ionic relaxation. A \(\Gamma\)-centered k-point grid of 2\(\pi\) × 0.02
\AA\textsuperscript{-1} spacing was used for the Brillouin zone sampling.
To calculate the magnonic dispersion, we used the Heisenberg model
approach with the spin-polarized version of RKKY exchange interaction
parameters \cite{41}. RKKY magnetic exchange coupling parameters were
computed by using the TB2J package \cite{42} based on localized orbitals
obtained by the OpenMX package \cite{43} with the PBE GGA functional.
Hubbard U was added using the Dudarev scheme \cite{44}. The magnonic dispersion 
was calculated by using the SpinW code \cite{45}. The crystal
orbital Hamilton populations (COHP) were calculated by using the LOBSTER
program \cite{46}.\\

\emph{Acknowledgments}---This work was supported by the U.S. Department of Energy (DOE)
Established Program to Stimulate Competitive Research (EPSCoR) Grant No.
DE-SC0024284. Computations were performed at the High Performance
Computing facility at Iowa State University and the Holland Computing
Center at the University of Nebraska.\\

\emph{Data availability}---The data that support the findings of this article are openly available.


\begin{thebibliography}{100}
\bibitem{20} Note that the tiny unphysical band spin splitting along the
nodal planes or lines in the charge-disproportionated band structure is
due to numerical errors in symmetry-unconstrained calculations, which is
negligible and can be reduced by increasing computational accuracy.
\end{thebibliography}
\end{document}